\documentclass{article}
\usepackage[utf8]{inputenc}
\usepackage{graphicx}
\usepackage{color}
\usepackage{multirow}
\usepackage{latexsym}
\usepackage{amsmath,array}
\usepackage{tabularx}

\title{``Mysteries'' of Modern Physics and the Fundamental Constants $c$, $h$, and $G$}

\author{W.M. Stuckey\thanks{Department of Physics, Elizabethtown College, Elizabethtown, PA 17022, USA} \, Timothy McDevitt\thanks{Department of Mathematical Sciences, Elizabethtown College, Elizabethtown, PA 17022, USA} \, and Michael Silberstein\thanks{Department of Philosophy, Elizabethtown College, Elizabethtown, PA 17022, USA} \, \thanks{Department of Philosophy, University of Maryland, College Park, MD 20742, USA}
}

\begin{document}

\maketitle
\sloppy

\begin{abstract}
We review how the kinematic structures of special relativity and quantum mechanics both stem from the relativity principle, i.e., ``no preferred reference frame'' (NPRF). Essentially, NPRF applied to the measurement of the speed of light $c$ gives the light postulate and leads to the geometry of Minkowski spacetime, while NPRF applied to the measurement of Planck's constant $h$ gives ``average-only'' projection and leads to the denumerable-dimensional Hilbert space of quantum mechanics. These kinematic structures contain the counterintuitive aspects (``mysteries'') of time dilation, length contraction, and quantum entanglement. In this essay, we extend the application of NPRF to the gravitational constant $G$ and show that it leads to the ``mystery'' of the contextuality of mass in general relativity. Thus, we see an underlying coherence and integrity in modern physics via its ``mysteries'' and the fundamental constants $c$, $h$, and $G$.
\end{abstract}

\clearpage

All undergraduate physics majors are shown how the counterintuitive aspects (``mysteries'') of time dilation and length contraction in special relativity (SR) follow from the light postulate, i.e., that everyone measures the same value for the speed of light $c$, regardless of their motion relative to the source. And, we can understand the light postulate to follow from the principle of relativity, sometimes referred to as ``no preferred reference frame'' (NPRF). Simply put, if the speed of light from a source was only equal to $c=\frac{1}{\sqrt{\epsilon_o \mu_o}}$ (per Maxwell's equations) for one particular velocity relative to the source, that would certainly constitute a preferred reference frame  \cite{serway,knight}.  

While time dilation and length contraction follow ``analytically'' from the light postulate, there are those who do not consider the light postulate explanatory, since it does not provide ``hypothetically constructed'' mechanisms to ``synthetically'' account for time dilation and length contraction \cite{brownbook,brownpooley2006}. That is, the postulates of SR are principles offered without corresponding ``constructive efforts.'' In what follows, Einstein explains the difference between the two \cite{einstein1919}:
\begin{quote}
We can distinguish various kinds of theories in physics. Most of them are constructive. They attempt to build up a picture of the more complex phenomena out of the materials of a relatively simple formal scheme from which they start out. Thus the kinetic theory of gases seeks to reduce mechanical, thermal, and diffusional processes to movements of molecules – i.e., to build them up out of the hypothesis of molecular motion. When we say that we have succeeded in understanding a group of natural processes, we invariably mean that a constructive theory has been found which covers the processes in question.\\

Along with this most important class of theories there exists a second, which I will call ``principle-theories.'' These employ the analytic, not the synthetic, method. The elements which form their basis and starting point are not hypothetically constructed but empirically discovered ones, general characteristics of natural processes, principles that give rise to mathematically formulated criteria which the separate processes or the theoretical representations of them have to satisfy. Thus the science of thermodynamics seeks by analytical means to deduce necessary conditions, which separate events have to satisfy, from the universally experienced fact that perpetual motion is impossible.\\

The advantages of the constructive theory are completeness, adaptability, and clearness, those of the principle theory are logical perfection and security of the foundations. The theory of relativity belongs to the latter class. In order to grasp its nature, one needs first of all to become acquainted with the principles on which it is based.
\end{quote}
Here is why Einstein formulated SR as a principle theory \cite[pp. 51-52]{einstein1949}:
\begin{quote}
By and by I despaired of the possibility of discovering the true laws by means of constructive efforts based on known facts. The longer and the more despairingly I tried, the more I came to the conviction that only the discovery of a universal formal principle could lead us to assured results.
\end{quote}
Despite the fact that ``there is no mention in relativity of exactly \textit{how} clocks slow, or \textit{why} meter sticks shrink'' (no ``constructive efforts''), the ``empirically discovered'' principles of SR are so compelling that ``physicists always seem so sure about the particular theory of Special Relativity, when so many others have been superseded in the meantime'' \cite{mainwood2018}. 

As it turns out, we are in a similar position today with quantum mechanics (QM). As emphasized by Fuchs, ``Where present-day quantum-foundation studies have stagnated in the stream of history is not so unlike where the physics of length contraction and time dilation stood before Einstein's 1905 paper on special relativity'' \cite{fuchs2002}. In other words, SR provides an historical precedent for dealing with QM. Moylan writes \cite{moylan2021}:
\begin{quote}
The point is that at the end of the nineteenth century, physics was in a terrible state of confusion. Maxwell's equations were not preserved under the Galilean transformations and most of the Maxwellian physicists of the time were ready to abandon the relativity of motion principle [3], [4]. They adopted a distinguished frame of reference which was the rest frame of the ``luminiferous aether,'' the medium in which electromagnetic waves propagate and in which Maxwell's equations and the Lorentz force law have their usual forms. In effect they were ready to uproot Copernicus and reinstate a new form of geocentricism.
\end{quote}
Even ``Einstein was willing to sacrifice the greatest success of 19th century physics, Maxwell’s theory, seeking to replace it by one conforming to an emission theory of light, as the classical, Galilean kinematics demanded'' before realizing that such an emission theory would not work \cite{norton2004}. 

QM is similar to SR in that it accurately predicts violations of the Clauser-Horne-Shimony-Holt (CHSH) inequality all the way to the Tsirelson bound for Bell state entanglement without providing a corresponding constructive account \cite{TsirelsonBound2019}. Violation of the CHSH inequality leads some to believe that QM and SR are fundamentally incompatible \cite{mamone}. Bell himself voiced concerns about the compatibility of SR and QM based on quantum entanglement \cite[p. 172]{bellbook}:
\begin{quote}
For me then this is the real problem with quantum theory: the apparently essential conflict between any sharp formulation and fundamental relativity. That is to say, we have an apparent incompatibility, at the deepest level, between the two fundamental pillars of contemporary theory.
\end{quote} 
That QM accurately predicts the violation of the CHSH inequality to the Tsirelson bound without spelling out any corresponding constructive account prompted Smolin to write \cite[p. xvii]{smolin}: 
\begin{quote}
I hope to convince you that the conceptual problems and raging disagreements that have bedeviled quantum mechanics since its inception are unsolved and unsolvable, for the simple reason that the theory is wrong. It is highly successful, but incomplete. 
\end{quote}
Of course, this is precisely the complaint leveled by Einstein, Podolsky, and Rosen (EPR) in their famous paper, ``Can Quantum-Mechanical Description of Physical Reality Be Considered Complete?'' \cite{EPRpaper}. 

The EPR paper was published in 1935 and yet physics still has no (consensus) constructive account of quantum entanglement. Thus, like Einstein with SR, physicists are starting to despair of finding a causal mechanism responsible for quantum entanglement. Hardy writes \cite[p. 224]{hardy2016}:
\begin{quote}
The standard axioms of [quantum theory] are rather ad hoc. Where does this structure come from? Can we write down natural axioms, principles, laws, or postulates from which we can derive this structure? Compare with the Lorentz transformations and Einstein's two postulates for special relativity. Or compare with Kepler's Laws and Newton's Laws. The standard axioms of quantum theory look rather ad hoc like the Lorentz transformations or Kepler's laws. Can we find a natural set of postulates for quantum theory that are akin to Einstein's or Newton's laws?
\end{quote}
Other physicists in quantum information theory are also calling for a principal account of QM. Fuchs writes \cite[p. 285]{fuchs1}:
\begin{quote}
Compare [quantum mechanics] to one of our other great physical theories, special relativity. One could make the statement of it in terms of some very crisp and clear physical principles: The speed of light is constant in all inertial frames, and the laws of physics are the same in all inertial frames. And it struck me that if we couldn't take the structure of quantum theory and change it from this very overt mathematical speak -- something that didn't look to have much physical content at all, in a way that anyone could identify with some kind of physical principle -- if we couldn't turn that into something like this, then the debate would go on forever and ever. And it seemed like a worthwhile exercise to try to reduce the mathematical structure of quantum mechanics to some crisp physical statements. 
\end{quote}
Along these lines, we recently showed that the qubit Hilbert space structure at the foundation of axiomatic reconstructions of quantum theory following from the information-theoretic principles of ``Existence of an Information Unit'' \cite{masanesMullerAugPerez2013} and ``Continuous Reversibility'' \cite{masanesMullerAugPerez2013}, or in combined form ``Information Invariance \& Continuity'' \cite{bruknerZeil2009}, entails NPRF applied to the invariant measurement of Planck's constant $h$ \cite{stuckey2021}. 

\begin{figure}
\begin{center}
\includegraphics [height = 40mm]{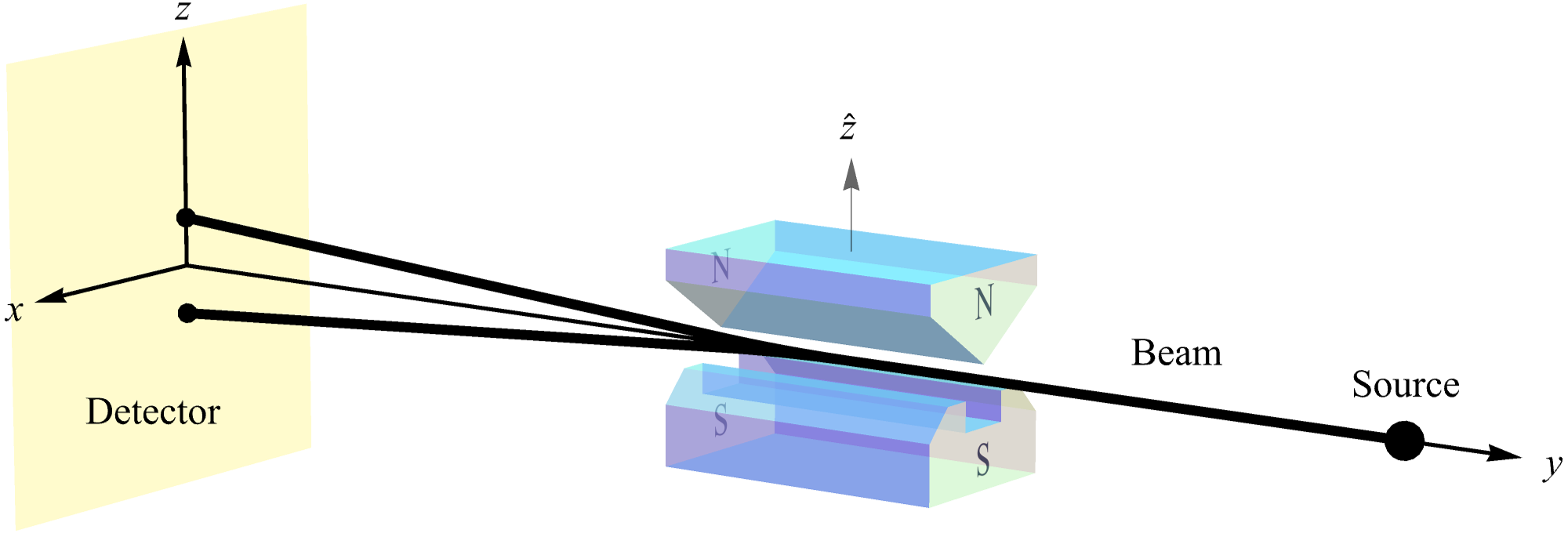}  \caption{A Stern-Gerlach (SG) spin measurement showing the two possible outcomes, up ($+\frac{\hbar}{2}$) and down ($-\frac{\hbar}{2}$) or $+1$ and $-1$, for short. As Weinberg points out, this constitutes a measurement of Planck's constant $h$.} \label{SGExp}
\end{center}
\end{figure}

As Weinberg points out, measuring an electron's spin via Stern-Gerlach (SG) magnets constitutes the measurement of ``a universal constant of nature, Planck's constant'' \cite[p. 3]{weinberg2017} (Figure \ref{SGExp}). So if NPRF applies equally here, everyone must measure the same value for Planck's constant \textit{h} regardless of their SG magnet orientations relative to the source, which like the light postulate is an ``empirically discovered'' fact. 

\begin{figure}
\begin{center}
\includegraphics [height = 75mm]{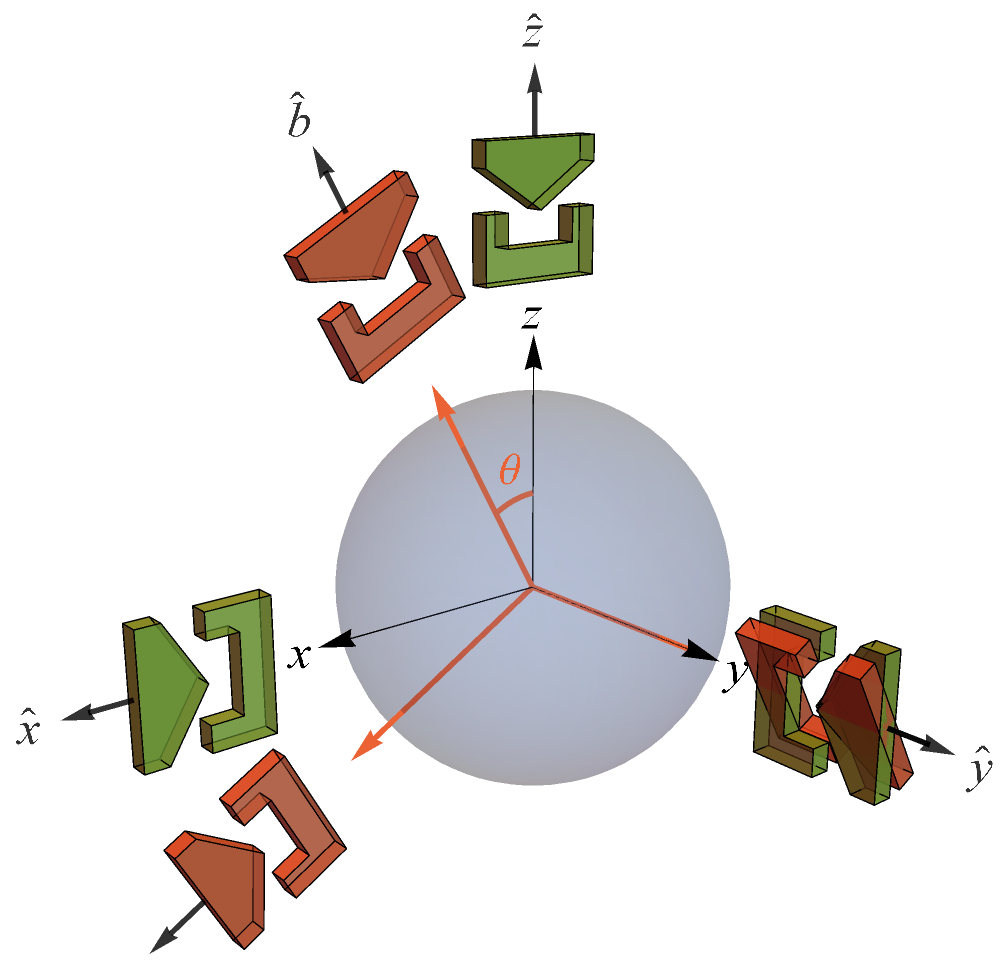} 
\caption{Two reference frames related by SO(3) each associated with a set of mutually complementary SG spin measurements \cite{bruknerZeil2003}.} \label{ComplBases}
\end{center}
\end{figure}

Here ``reference frame'' refers to a set of mutually complementary qubit measurements \cite{bruknerZeil2003} and these reference frames are related by SO(3) (Figure \ref{ComplBases}). More specifically, different 2-dimenionsal Hilbert space measurement operators with the same outcomes are related by SU(2) transformations and SU(2) transformations in Hilbert space map to SO(3) rotations between different reference frames in 3-dimensional real space (Information Invariance \& Continuity). Since SO(3) is a subgroup of both Lorentz and Galilean transformations between inertial reference frames, we see that this application of the relativity principle does not entail Lorentz invariance. Axiomatic reconstructions of denumerable-dimensional QM are built in composite fashion from this qubit structure \cite{stuckey2021}, thus the relativity principle is at the foundation of QM via the invariant measurement of Planck's constant $h$ exactly as it at the foundation of SR via the invariant measurement of the speed of light $c$. This leads to ``average-only'' projection responsible for the ``mystery'' of entanglement's ``average-only'' conservation.

\begin{figure}
\begin{center}
\includegraphics [height = 50mm]{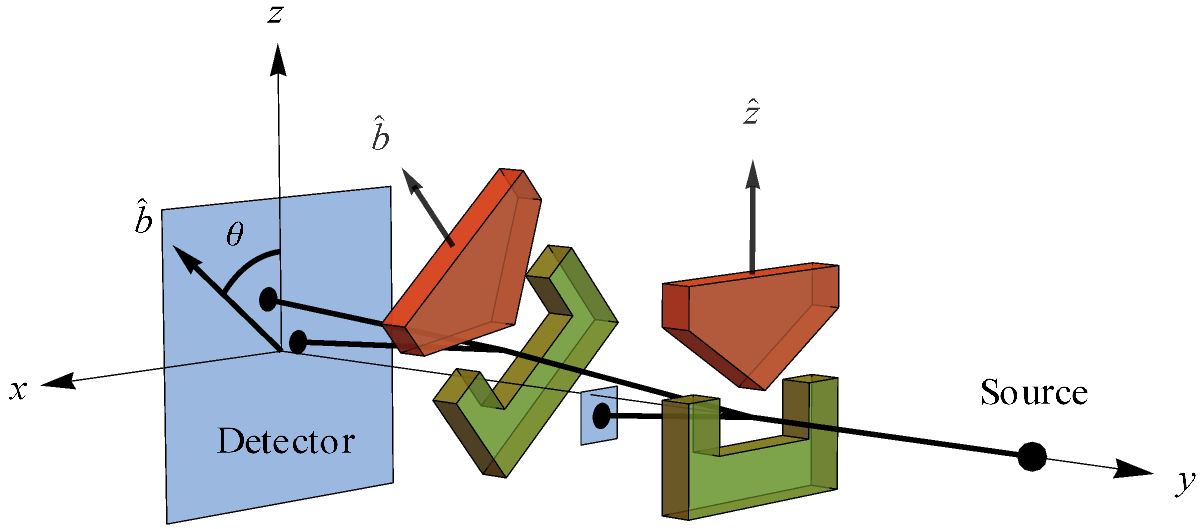}  
\caption{In this set up, the first SG magnets (oriented at $\hat{z}$) are being used to produce an initial state $|\psi\rangle = |u\rangle$ for measurement by the second SG magnets (oriented at $\hat{b}$). } \label{SGExp2}
\end{center}
\end{figure}

\begin{figure}
\begin{center}
\includegraphics [height = 75mm]{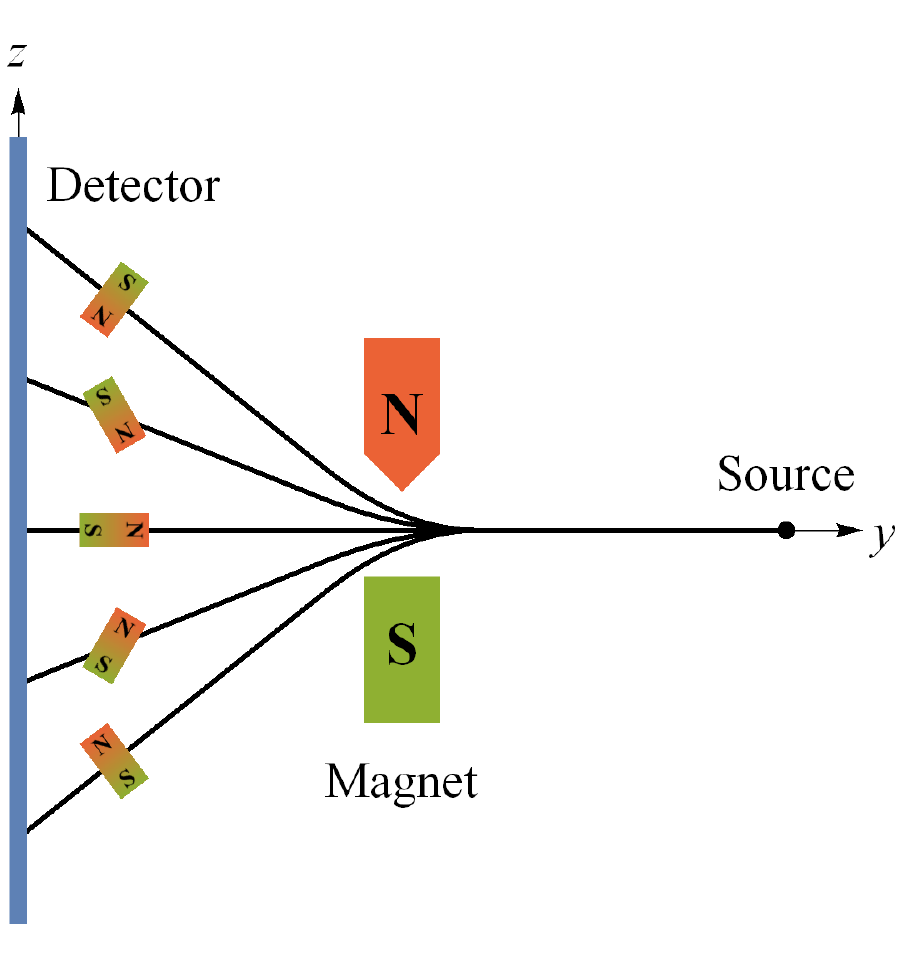}
\caption{The classical constructive model of the Stern-Gerlach (SG) experiment. If the atoms enter with random orientations of their ``intrinsic'' magnetic moments, the SG magnets should produce all possible deflections, not just the two that are observed \cite{knight,franklin2019}.} \label{SGclassical}
\end{center}
\end{figure}

\begin{figure}
\begin{center}
\includegraphics [height = 65mm]{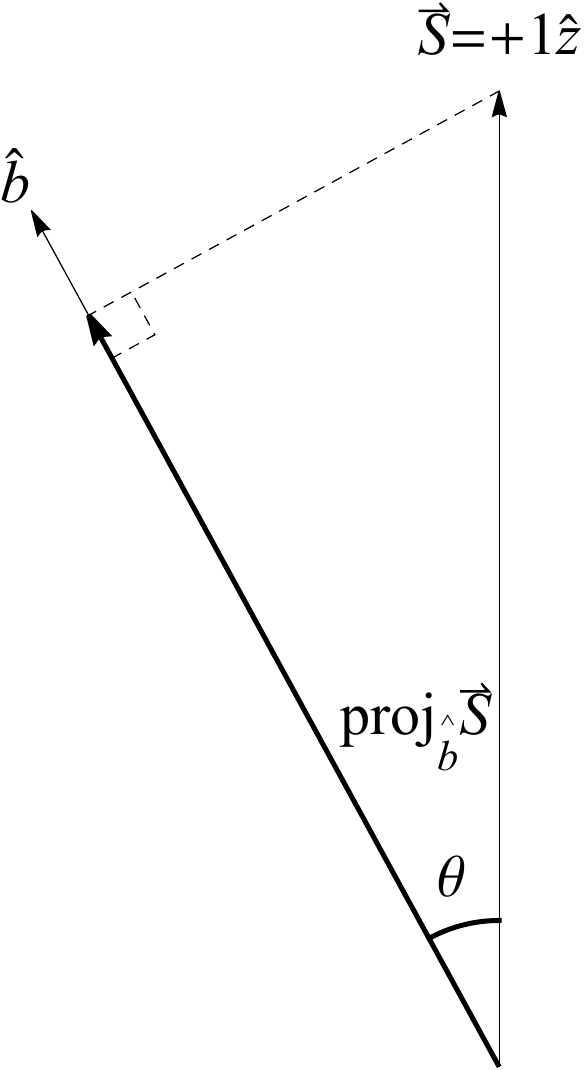} 
\caption{The ``intrinsic'' angular momentum of Bob's particle $\vec{S}$ projected along his measurement direction $\hat{b}$. This does \textit{not} happen with spin angular momentum due to NPRF.} \label{Projection}
\end{center}
\end{figure}

To see that, create a preparation state oriented along the positive $z$ axis as in Figure \ref{SGExp2}, i.e., $|\psi\rangle = |u\rangle$, so that our ``intrinsic'' angular momentum is $\vec{S} = +1\hat{z}$ (in units of $\frac{\hbar}{2} = 1$). Now proceed to make a measurement with the SG magnets oriented at $\hat{b}$ making an angle $\theta$ with respect to $\hat{z}$ (Figure \ref{SGExp2}). According to the constructive account of classical physics \cite{knight,franklin2019} (Figure \ref{SGclassical}), we expect to measure $\vec{S}\cdot\hat{b} = \cos{(\theta)}$ (Figure \ref{Projection}), but we cannot measure anything other than $\pm 1$ due to NPRF (contra the prediction by classical physics). As a consequence, we can only recover $\cos{(\theta)}$ \textit{on average}, i.e., NPRF dictates ``average-only'' projection
\begin{equation}
(+1) P(+1 \mid \theta) + (-1) P(-1 \mid \theta) = \cos (\theta) \label{AvgProjection}
\end{equation}
Of course, this is precisely $\langle\sigma\rangle$ per QM. Eq. (\ref{AvgProjection}) with our normalization condition \linebreak $P(+1 \mid \theta) +  P(-1 \mid \theta) = 1$ then gives 
\begin{equation}
P(+1 \mid \theta) = \mbox{cos}^2 \left(\frac{\theta}{2} \right) \label{UPprobability}
\end{equation}
and 
\begin{equation}
P(-1 \mid \theta) = \mbox{sin}^2 \left(\frac{\theta}{2} \right) \label{DOWNprobability}
\end{equation} 
again, precisely in accord with the qubit Hilbert space structure of QM. And, if we identify the preparation state $|\psi\rangle = |u\rangle$ at $\hat{z}$ with the reference frame of mutually complementary spin measurements $[J_x,J_y,J_z]$, then the SG measurement at $\hat{b}$ constitutes a reference frame of mutually complementary measurements rotated by $\theta$ in real space relative to the preparation state (Figure \ref{ComplBases}). Thus, ``average-only'' projection follows from Information Invariance \& Continuity when applied to SG measurements in real space. This explains the ineluctably probabilistic nature of QM, as pointed out by Mermin \cite[p. 10]{mermin2019}:
\begin{quote}
Quantum mechanics is, after all, the first physical theory in which probability is explicitly not a way of dealing with ignorance of the precise values of existing quantities.
\end{quote}
Consequently, QM is as ``complete as possible'' given NPRF \cite{silberstein2021}. This extends to ``average-only'' conservation for a pair of spin entangled particles responsible for the ``mystery'' of entanglement.

\begin{figure}
\begin{center}
\includegraphics [height = 70mm]{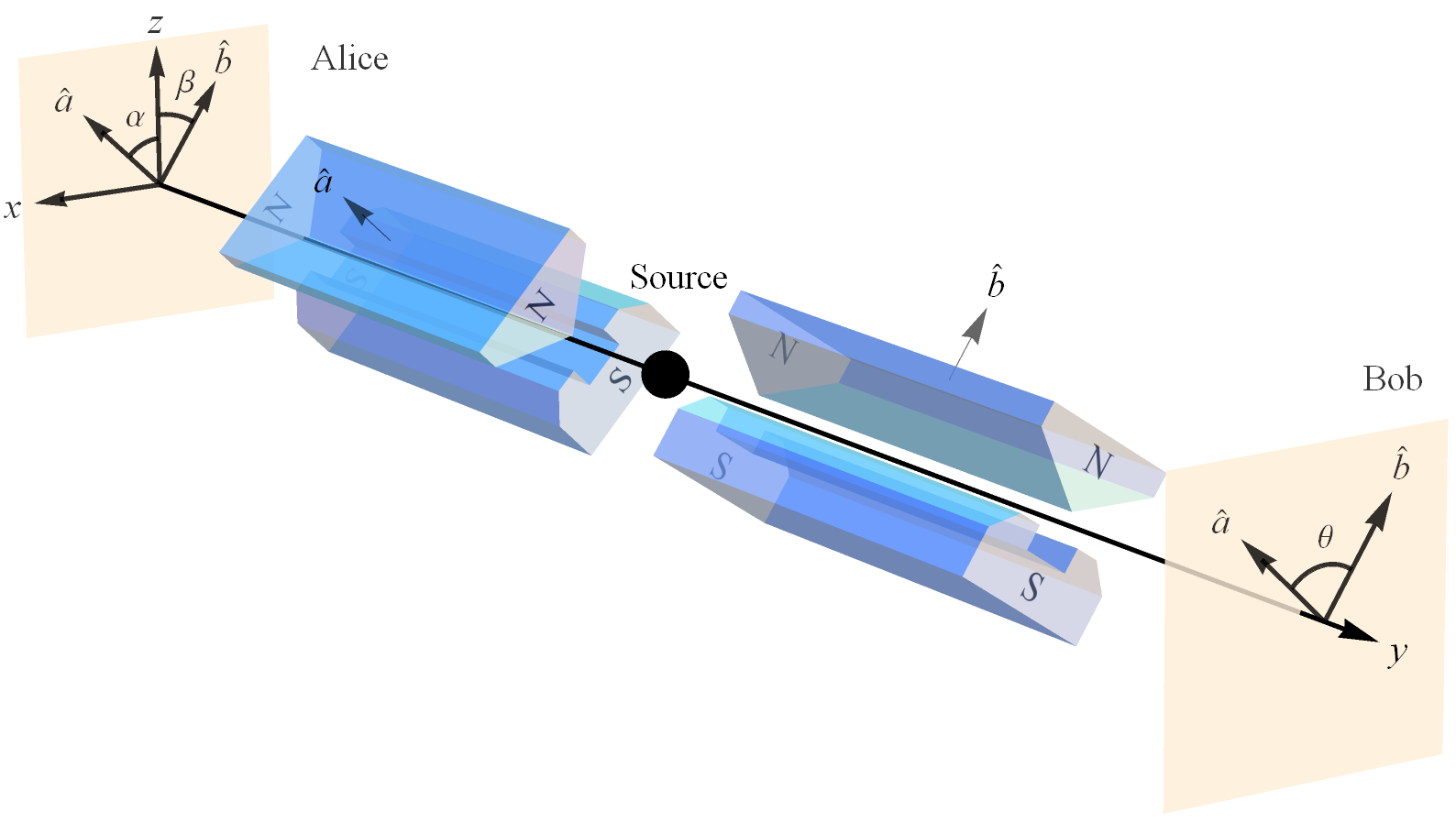}  \caption{Alice and Bob making spin measurements on a pair of spin-entangled particles in a Bell state with their SG magnets and detectors.} \label{EPRBmeasure}
\end{center}
\end{figure}

\begin{figure}
\begin{center}
\includegraphics [width=\textwidth]{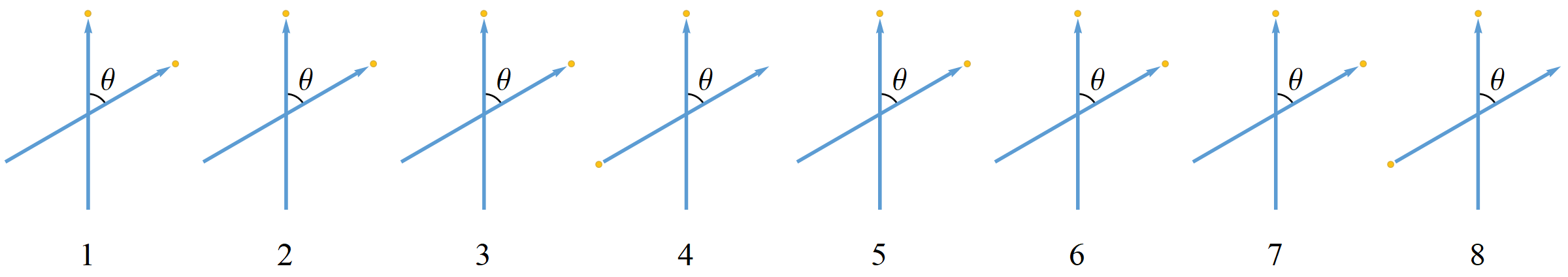} 
\caption{A spatiotemporal ensemble of 8 experimental trials for the spin triplet states showing Bob’s outcomes corresponding to Alice's $+1$ outcome when $\theta = 60^\circ$. Blue arrows depict SG magnet orientations and yellow dots depict the measurement outcomes. Spin angular momentum is not conserved in any given trial, because there are two different measurements being made, i.e., outcomes are in two different reference frames, but it is conserved on average for all 8 trials (six up outcomes and two down outcomes average to $\cos{(60^\circ)}=\frac{1}{2}$).} \label{4Dpattern}
\end{center}
\end{figure}

\begin{figure}
\begin{center}
\includegraphics [height = 18mm]{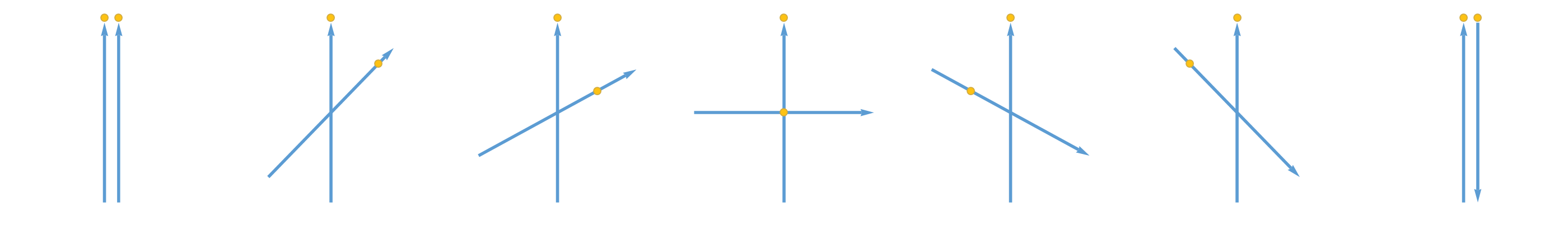}  \caption{\textbf{Average View for the Spin Triplet States}. Reading from left to right, as Bob rotates his SG magnets relative to Alice's SG magnets for her $+1$ outcome, the average value of his outcome varies from $+1$ (totally up, arrow tip) to $0$ to $-1$ (totally down, arrow bottom). This obtains per conservation of spin angular momentum on average in accord with no preferred reference frame. Bob can say exactly the same about Alice's outcomes as she rotates her SG magnets relative to his SG magnets for his $+1$ outcome. That is, their outcomes can only satisfy conservation of spin angular momentum on average in different reference frames, because they only measure $\pm 1$, never a fractional result. Again, just as with the light postulate of special relativity, we see that no preferred reference frame leads to a counterintuitive result. Here it requires quantum outcomes $\pm 1 \left(\frac{\hbar}{2}\right)$ for all measurements leading to the ``mystery'' of ``average-only'' conservation.} \label{AvgViewTriplet}
\end{center}
\end{figure}

\begin{figure}
\begin{center}
\includegraphics [height = 42mm]{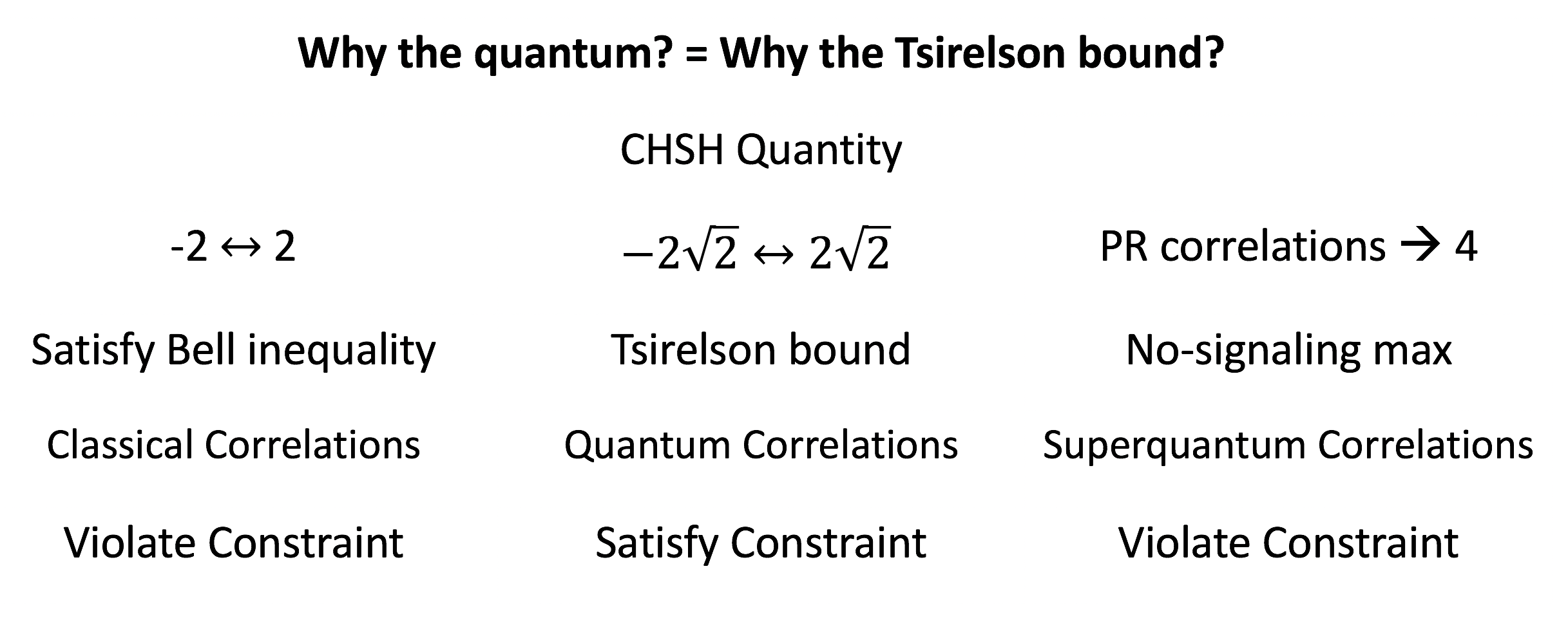}  \caption{Bub's version of Wheeler's question ``Why the quantum?'' is ``Why the Tsirelson bound?'' \cite{bub2004,bub2012}. The ``constraint'' is conservation per no preferred reference frame.} \label{Summary}
\end{center}
\end{figure}

\begin{figure}
\begin{center}
\includegraphics [height = 30mm]{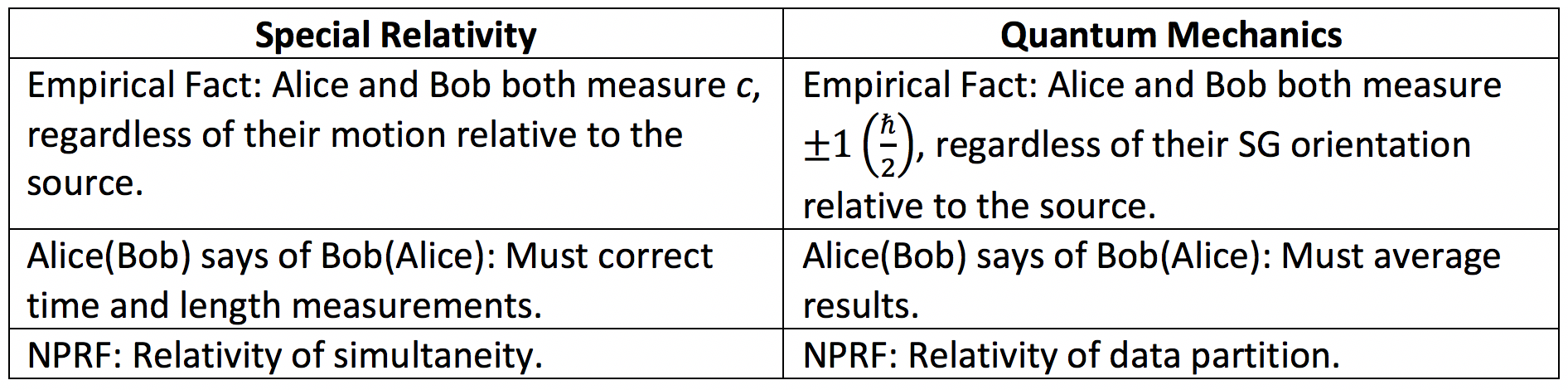}  \caption{\textbf{Comparing SR with QM according to no preferred reference frame (NPRF)}. Because Alice and Bob both measure the same speed of light $c$, regardless of their motion relative to the source per NPRF, Alice(Bob) may claim that Bob's(Alice's) length and time measurements are erroneous and need to be corrected (length contraction and time dilation). Likewise, because Alice and Bob both measure the same values for spin angular momentum $\pm 1$ $\left(\frac{\hbar}{2}\right)$, regardless of their SG magnet orientation relative to the source per NPRF, Alice(Bob) may claim that Bob's(Alice's) individual $\pm 1$ values are erroneous and need to be corrected (averaged, Figures \ref{4Dpattern} \& \ref{AvgViewTriplet}). In both cases, NPRF resolves the ``mystery'' it creates. In SR, the apparently inconsistent results can be reconciled via the relativity of simultaneity. That is, Alice and Bob each partition spacetime per their own equivalence relations (per their own reference frames), so that equivalence classes are their own surfaces of simultaneity and these partitions are equally valid per NPRF. This is completely analogous to QM, where the apparently inconsistent results per the Bell spin states arising because of NPRF can be reconciled by NPRF via the ``relativity of data partition.'' That is, Alice and Bob each partition the data per their own equivalence relations (per their own reference frames), so that equivalence classes are their own $+1$ and $-1$ data events and these partitions are equally valid.} \label{SRvQM}
\end{center}
\end{figure}

Specifically, when Alice and Bob make their SG spin measurements on a Bell spin state (Figure \ref{EPRBmeasure}) at the same angle in the plane of symmetry (same reference frame), conservation of spin angular momentum dictates that they obtain the same result (both $+1$ or both $-1$) for the spin triplet states (opposite results for the spin singlet state) \cite{MerminChallenge2020}. Thus again, classical physics suggests that if Bob makes his SG spin measurement at angle $\theta$ with respect to Alice (different reference frames), then according to Alice he should obtain $\cos{(\theta)}$ when she obtains $+1$ in accord with the conservation of ``intrinsic'' angular momentum. But, Bob can only ever measure $\pm 1$ per the relativity principle, just like Alice, so the conservation principle is constrained to hold only \textit{on average} per NPRF (Figures \ref{4Dpattern} \& \ref{AvgViewTriplet}). Thus, Bell state entanglement and the Tsirelson bound are ``mysteries'' precisely because of ``average-only'' conservation, which follows from ``average-only'' projection per NPRF, so this is conservation per NPRF (Figure \ref{Summary}). Consequently, we see that the relativity principle reveals an underlying coherence between (non-relativistic) QM and SR (Figure \ref{SRvQM}) where others have perceived tension \cite{bellbook,mamone}. 

Of course, we know QM is not Lorentz invariant and so it deviates trivially from SR in that fashion. In order to get QM from Lorentz invariant quantum field theory one needs to make low energy approximations \cite[p. 173]{zee}. But, the charge of incompatibility based on QM entanglement actually carries serious consequences, because we have experimental evidence confirming the violation of the CHSH inequality per QM entanglement. So, if the violation of the CHSH inequality is in any way inconsistent with SR, then SR is being challenged empirically. By analogy, we know Newtonian mechanics deviates from SR because it is not Lorentz invariant. As a consequence, Newtonian mechanics predicts a very different velocity addition rule, so suppose we found experimentally that velocities do add as predicted by Newtonian mechanics. That would not merely mean that Newtonian mechanics and SR are incompatible, that would mean Newtonian mechanics has been empirically verified while SR has been empirically refuted. So, if one believes the violation of the CHSH inequality is in any way inconsistent with SR, and one believes the experimental evidence is accurate, then one believes SR has been empirically refuted. Clearly that is not the case, so their reconciliation as regards the violation of the CHSH inequality must certainly obtain in some fashion and here we see how the principle of NPRF does the job. Further, contrary to Smolin and EPR, Bell state entanglement does not mean that QM is ``incomplete'' or ``wrong.'' Rather, QM is as complete as possible per NPRF.

Given this result, one immediately wonders if general relativity (GR) can be brought into the mix via NPRF and the gravitational constant $G$. Of course it can and the associated counterintuitve aspect (``mystery'') in GR is the contextuality of mass. We have already shown how this might resolve the missing mass problem without having to invoke non-baryonic dark matter \cite{stuckeyDM,stuckeyDM2}.  

Specifically, we are pointing out the well-known result per GR that matter can simultaneously possess different values of mass when it is responsible for different combined spatiotemporal geometries. Here ``reference frame'' refers to each of the different spatiotemporal geometries associated with one and the same matter source. Tacitly assumed in this result is of course that $G$ has the same value in each reference frame, which is consistent with NPRF a la $c$ and $h$ above. This spatiotemporal contextuality of mass is not present in Newtonian gravity where mass is an intrinsic property of matter. For example, when a Schwarzschild vacuum surrounds a spherical matter distribution the ``proper mass'' $M_{p}$ of the matter, as measured locally in the matter, can be different than the ``dynamic mass'' $M$ in the Schwarzschild metric responsible for orbital kinematics about the matter \cite[p. 126]{wald}. This difference is attributed to binding energy and goes as $\displaystyle dM_p = \left(1-\frac{2GM(r)}{c^2r}\right)^{-1/2} \: dM$. In another example, suppose a Schwarzschild vacuum surrounds a sphere of Friedmann-Lema{\^i}tre-Robertson-Walker (FLRW) dust, as used originally to model stellar collapse \cite[pp. 851-853]{misner}. The dynamic mass $M$ of the surrounding Schwarzschild metric is related to the proper mass $M_{p}$ of the FLRW dust, as joined at FLRW radial coordinate $\chi_o$, by
\begin{equation}
\frac{M_p}{M} = \frac{3(2\chi_o -\sin(2\chi_o))}{4 \sin ^3(\chi_o)} \label{massratio}
\end{equation}
where
\begin{equation}
ds^2 = -c^2d\tau^2 + a^2(\tau)\left(d\chi^2 + \sin^2\chi d\Omega^2 \right) \label{FLRWmetric}
\end{equation}
is the closed FLRW metric \cite{stuckeyAJP4}. We should quickly point out that this may prima facie seem to constitute a violation of the equivalence principle, as understood to mean inertial mass equals gravitational mass, since inertial mass can't be equal to two different values of gravitational mass. But, the equivalence principle says simply that spacetime is locally flat \cite[pp. 68-69]{weinberg} and that is certainly not being violated here nor with any solution to Einstein's equations. 

Thus, contrary to what many believe about SR, QM, and GR collectively, these theories are comprehensive (not ``incomplete'' per \cite{smolin} and \cite{EPRpaper}) and coherent (not ``in conflict'' per \cite{mamone} and \cite{bellbook}). In order to appreciate the beauty of these theories collectively, one need only view them per the relativity principle (NPRF) with their associated ``mysteries'' corresponding to $c$, $h$, and $G$, respectively.

\newpage

\clearpage

\bibliography{biblio.bib}
\bibliographystyle{spphys}
\end{document}